\documentclass[fleqn]{aa}
\usepackage[varg]{txfonts}
\usepackage{graphicx, subfigure}
\usepackage{natbib}
\usepackage{verbatim}
\usepackage{txfonts,epsfig,url,twoopt}
\usepackage[breaklinks=true]{hyperref}
\hypersetup{colorlinks=true,citecolor=blue}
\bibpunct{(}{)}{;}{a}{}{,} 
\usepackage{color} 
\usepackage{cleveref}
\usepackage{nicefrac}

\begin{document}
	\title{The role of disc torques in forming resonant planetary systems}	

	\titlerunning{Tighter resonances with inconsistent torques}
	\authorrunning{S.~Ataiee \& W.~Kley}
	
	\author{S.~Ataiee\thanks{sareh.ataiee@uni-tuebingen.de}
		\and 
		W.~Kley}
	\institute{Institut f\"ur Astronomie \& Astrophysik, Universit\"at T\"ubingen, Auf der Morgenstelle 10, 72076 T\"ubingen, Germany}
	
	\abstract{The most accurate method for modelling planetary migration and hence the formation of resonant systems is using hydrodynamical simulations. Usually, the force (torque) acting on a planet is calculated using the forces from the gas disc and the star, while the gas accelerations are computed using the pressure gradient, the star, and the planet's gravity, ignoring its own gravity. For the non-migrating the neglect of the disc gravity results in a consistent torque calculation while for the migrating case it is inconsistent.}
	{We aim to study how much this inconsistent torque calculation can affect the final configuration of a two-planet system. Our focus will be on low-mass planets because most of the multi-planetary systems, discovered by the \textit{Kepler} survey, have masses around 10~Earth masses.}
	{Performing hydrodynamical simulations of planet-disc interaction, we measure the torques on non-migrating and migrating planets for various disc masses as well as density and temperature slopes with and without considering the disc self-gravity. Using this data, we find a relation that quantifies the inconsistency, use it in an N-body code, and  perform an extended parameter study modelling the migration of a planetary system with different planet mass ratios and disc surface densities, in order to investigate the impact of the torque inconsistency on the architecture of the planetary system.}
	{Not considering disc self-gravity produces an artificially larger torque on the migrating planet that can result in tighter planetary systems. The deviation of this torque from the correct value is larger in discs with steeper surface density profiles. 
   }
	{In hydrodynamical modelling of multi-planetary systems, it is crucial to account for the torque correction, otherwise the results favour more packed systems. We examine two simple correction methods existing in the literature and show that they properly correct this problem.}
	
	\keywords{Hydrodynamics - Methods: numerical-Planetary systems - Protoplanetary disks - Planet-disk interactions}
	
	\maketitle
	\section{Introduction}
	\label{sec:intro}
	Among thousands of known exo-planets, hundreds of them are in multi-planetary systems, mostly discovered in the \textit{Kepler} survey. One of the well-known properties of \textit{Kepler} multi-planetary systems is the existence of pile-ups in the orbital period-ratio distribution just wide of some commensurabilities, specifically 2:1 and 3:2 \citep[e.g.][]{2014ApJ...790..146F}. Moreover, there are systems which have been shown to reside in resonance chains, such as Trappist-1 \citep{2017NatAs...1E.129L}, GJ~876 \citep{2018AJ....155..106M}, Kepler-223 \citep{2016Natur.533..509M}, and HD~40307 \citep{2009A&A...493..639M}. These observations demonstrate that there should be one or more mechanisms that place the planets into these fine-tuned configurations. One of the most promising mechanism that is able to arrange planets in resonant configurations is planetary migration --the drift of a planet as a result of its gravitational interaction with the natal disc. There are numerous studies on modelling resonance configurations, pioneered by \cite{1993Natur.365..819M} who shows that the migration of planets, due to their interaction with either gas or planetesimal disc, can efficiently bring the planets into resonance. Forming resonant configurations requires convergent migration, meaning the outer planet should have larger migration rate than the inner one, assuming the migration of both is towards the star. In convergent migration, the outer planet migrates faster, catches the inner one in a resonance, and then they continue their migration maintaining their orbital configuration \citep[e.g.][]{2002ApJ...567..596L,2001A&A...374.1092S}. The final resonance configuration depends on how fast the outer planet crosses the resonance locations. The faster the outer planet migrates towards the inner one, the closer the resonance configuration will be. Therefore, in modelling of such systems, the migration speed of the planets is a key parameter.
		
	Migration of a planet in a gas disc is determined by angular momentum exchange with the disc at the location of Lindblad and co-rotation resonances \citep[see ][for a review]{2012ARA&A..50..211K}. Ignoring the gas pressure and self-gravity, the resonances are located at those radii where the rotational frequency of the gas relative to the planet, $\Omega(r)-\Omega(r_{\rm p})$, matches a multiple of the epicyclic frequency in the disc $\kappa(r)$. In other words, resonances are where the relation $m(\Omega(r)-\Omega(r_{\rm p}))=\pm \kappa(r)$, with $m$ being an integer, is satisfied. Whatever affects the gas angular frequency shifts the location of resonances and consequently changes the torque.

	The angular velocity of a planet in a typical proto-planetary disc, which it is not so massive and cold to be prone to gravitational instability, is dictated by the gravity from the central star and the disc. The angular velocity of the gas in the absence of the planet, $\Omega(r)$, is determined by the gravity of the star, the disc's pressure gradient, and also the disc's own gravity. In hydrodynamical modelling of planet-disc interaction, the third contribution, the disc's self-gravity is usually ignored for moderate disc masses of about minimum mass solar nebula (MMSN) in order to save the computation time. However, by means of analytical calculation \citep{2005A&A...433L..37P} and numerical simulations \citep{2008ApJ...678..483B} it was shown that the torque on the planet would be more negative if the calculations of planet and gas angular velocities are \textit{inconsistent}. This inconsistency occurs when the migrating planet planet feels both the star and the disc but the gas only feels the gravity of the star and not its own. In such a condition, the term $\Omega(r)-\Omega(r_{\rm p})$ differs from the correct physical condition, when the disc self-gravity is included. This way of hydrodynamical simulations with  migrating planets in a non-self-gravitating disc is very common in the literature \citep[e.g.][]{2001A&A...374.1092S,2004A&A...414..735K,2005MNRAS.363..153P,2011MNRAS.417.2253P}.
	
	The reason is that it is not expected that a slightly faster migration affects the outcome significantly. However, in our attempt of modelling resonant planets with hydrodynamical simulations, we noticed that neglecting this effect produces unexpected resonant configurations and therefore, using the consistent calculation of the torque is essential. One should note that the torques obtained from hydrodynamical simulations using a non-migrating planet in a non-self-gravitating disc are indeed consistent because the planet's angular velocity is enforced to be Keplerian. It means the planet only feels the star \citep[e.g.][]{2010MNRAS.401.1950P,2011MNRAS.410..293P}. Therefore, these torques, which are widely used in one-dimensional~(1D) or N-body models \citep[e.g.][]{2014A&A...567A.121D,2015A&A...575A..28B,2018ApJ...864L...8B}, are consistent.
	
	There are several ways to obtain  a consistent calculation of the torque for migrating planets while avoiding costly full self-gravitating simulations. One method, which is suggested by \cite{2008ApJ...678..483B} and applied by \cite{2016ApJ...826...13B}, is to use only the perturbed surface density for calculating the torque on the planet. This method removes the acceleration from the whole disc on the planet except for the non-axisymmetric perturbations induced by the planets such as spirals. Another method, also suggested by \cite{2008ApJ...678..483B}, is including the axi-symmetric part of the disc self-gravity in the calculation, assuming the contribution of the non-axi-symmetric part in the velocity of gas is negligible.
	
	In this study we investigate how much an inconsistent torque calculation impacts the final configuration of a two-planet system. In Sec.~\ref{sec:torquesingle}, we initially examine how much the inconsistent torque differs from the correct torque for discs with various surface density and temperature profiles. Then we examine two above-cited correction methods and show that both perfectly fix the torque. Afterwards, in Sec.~\ref{sec:twophydro}, we present the results of hydrodynamical simulations with two planets that highlight the importance of torque correction and show that these two methods give similar results to their full self-gravitating counterparts. In Sec.~\ref{sec:parameterstudy}, we first present a relation that gives the ratio of the correct torque to the inconsistent torque. This relation helps us to perform a parameter study using an N-body code for comparing the outcome of two-planet simulations with the correct and the inconsistent toque. Finally, we summarise our results in Sec.~\ref{sec:summary}.
	
	\section{Migration of a single planet}
	\label{sec:torquesingle}
		In this section, we will examine how much the torque on a migrating low-mass planet changes if we ignore the effect of disc self-gravity. We present this torque differences for various disc profiles and planetary masses, as well as a comparison to the widely used torque formula of \cite{2011MNRAS.410..293P} (hereafter P11). These results will be used later in Sec.~\ref{sec:parameterstudy} in the N-body simulations. Then we will introduce and apply two correction methods, which already existed in the literature and mentioned in Sec.~\ref{sec:intro}, on the migrating torque and compare the outcome with the non-migrating and full self-gravitating models.
		
\begin{figure}
			\centerline{\includegraphics[width=\columnwidth]{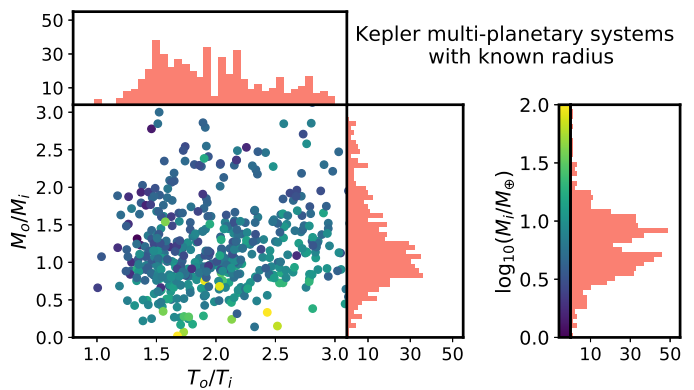}}
			\caption{Mass and orbital period ratios for two adjacent planets of \textit{Kepler} multi-planetary systems. For the planets with unmeasured masses, we used the mass-radius relation $M_{p}/M_{\oplus} = 2.7 (R_{p}/R_{\oplus})^{1.3}$ by \cite{2016ApJ...825...19W}, where $M_{p}/M_{\oplus}$ and $R_\mathrm{p}/R_{\oplus}$ are the planet-to-Earth mass and radius ratios. The subscripts $o$ and $i$ refer to the quantity of \textit{outer} and \textit{inner} planet in each pair. The colour of every point represents the mass of inner planet. For each quantity, we also attached a corresponding histogram to show its overall distribution. As the two mass-related histograms show, the masses of inner planets are between 5--20$M_{\oplus}$ while the mass ratio distribution peaks around unity.}
			\label{fig:keplermasses}
		\end{figure}

		In this study, we focus only on low-mass planets, which do not perturb the disc greatly (i.e. do not open a gap), because most of the multi-planetary systems discovered by the \textit{Kepler} space mission have Earth to mini-Neptune sizes. Using an exo-planet database\footnote{We used http://www.openexoplanetcatalogue.com/} and the mass-radius relation for low-mass planets by \cite{2016ApJ...825...19W}, we found that most of these planets have masses about 5--20$M_{\oplus}$ that can indeed be considered low-mass for a typical disc around a solar-type star. Figure~\ref{fig:keplermasses} shows the distribution of mass versus orbital period ratio for each adjacent pair in \textit{Kepler} multi-planet systems where the colour represents the mass of the inner planet. As the attached histogram to the colourbar shows, the mass of these planets is mostly between 5 to 20$M_{\oplus}$.
		
		\subsection{Method and numerics}
		\label{subsec:numericshydro}
				In order to measure the difference between migrating and non-migrating torques, we perform locally isothermal hydrodynamical simulations using the \texttt{FARGO-ADSG} code\footnote{http://fargo.in2p3.fr/-Legacy-archive-} \citep{2008ApJ...678..483B}. The disc surface density and temperature radial profiles follow power-lows $\Sigma=\Sigma_{0} (r/r_0)^{-\alpha}$ and $T=T_{0}(r/r_{0})^{-\beta}$, where $r_{0}$ is the unit of length, chosen here to be $1$au. The temperature profile is related to the disc aspect ratio $h$ through the sound speed ($c_\mathrm{s} \propto \sqrt{T}$) as $h=h_{0}(r/r_0)^f = c_\mathrm{s}/v_\mathrm{K}$ where $v_\mathrm{K}$ is Keplerian velocity, $f$ is flaring index, and $ h_{0}=0.05$. Therefore, $\beta=-2f+1$. We also respect the viscous equilibrium of the system by choosing the initial condition such that the mass accretion through the disc is constant $\dot{M}=3\pi\nu\Sigma$. This imposes the condition  $\alpha=2f+1/2$ for an alpha-viscosity model $\nu=\alpha_{\nu} c_\mathrm{s} H$, that is used in this study. This constant $\dot{M}$ condition is important when a model needs to be simulated for a long time such that the viscous evolution of the disc might pollute the results otherwise. The disc, that is spanned from $r=0.3$ to $2.5$ and $\phi$ over the whole $2\pi$, is divided by $\rm N_{r} \times N_{s} = 512 \times 1024$ grid cells. The spacing is logarithmic in radial and equidistant in azimuthal direction. This resolution is sufficient for resolving the horse-shoe region of planets with 10~Earth-masses and larger. In cases with smaller planets, we have increased the resolution correspondingly. However, we did not find any considerable change in the results if this resolution is used for smaller planetary masses down to 3 Earth-mass. The non-reflecting boundary condition is applied on the radial direction to damp the waves from the planet. The planet's smoothing length for the gravitational potential is $0.6 H$ where $H$ is the disc scale height. 
				
				In those simulations which include the axi-symmetric part of the self-gravity or the disc's full self-gravity, we consider the self-gravity smoothing length $\epsilon_{SG}$ equals to 0.6.

			\begin{figure*}
				\centerline{\includegraphics[width=\textwidth]{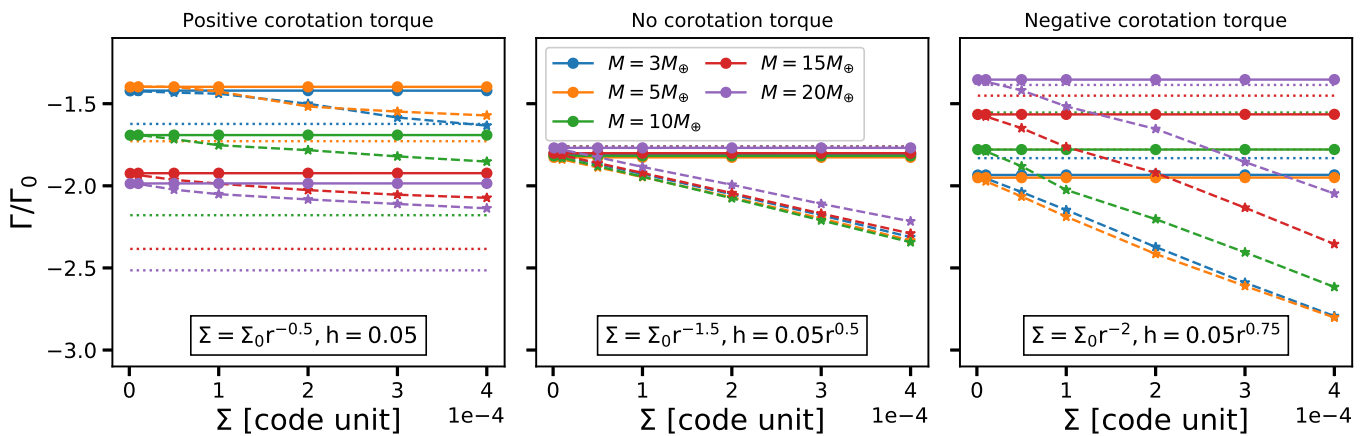}}
				\caption{The scaled torque on a migrating (stars/dashed lines) and non-migrating (bullets/solid lines) planet as a function of surface density, $\Sigma_0$ and planet mass $M$. The dotted lines mark the partially saturated (P11) torques. These plots show the importance of disc self-gravity correction even for moderate surface density values. We continued the simulations with non-zero vortensity gradient long enough to have the co-rotation torque established. }
				\label{fig:singlediscprofiles}
			\end{figure*}

	    \subsection{Migrating vs non-migrating torque}
	    \label{subsec:lowmass}
            In this section we study the torque acting on an embedded planet, non-migrating and migrating, for the case where no disc self-gravity is considered. In the first step, we use for the density and temperature slopes the parameters $\alpha=1.5$ and $f = 0.5$, that correspond to a locally isothermal disc with constant background vortensity. It guarantees that the torque on the planet is only generated by the Lindblad torque and avoids the complexity of the co-rotation torque \citep[e.g.][]{2006ApJ...642..478M,2008A&A...485..877P}.
	    	In the next step, we alter the disc surface density and temperature profiles such that the gradient of inverse background vortensity becomes positive or negative, and repeat the simulations again.

			The planet mass is varied between $\rm [3-20]M_{\oplus}$ where $\rm M_{\oplus}=3\times 10^{-6} M_{\star}$ is an Earth mass if $\rm M_{\star}$, which is the mass of central star and also mass unit in our simulations, equals to a solar mass. In order to see how much an inconsistent calculation (i.e. neglecting the effect of disc self-gravity) can alter the torque, we ran different models with various disc surface densities $\Sigma_{0}=[10^{-6}, 10^{-5}, 10^{-4}, 2\times 10^{-4}, 3\times 10^{-4}, 3\times 10^{-4}]$, and compared the torque of a non-migrating planet to an identical simulation where the planet is allowed to migrate. The value of $\Sigma_{0}=2 \times 10^{-4}$ corresponds to the surface density at $1$au in the MMSN. The viscosity parameter is $\alpha_{\nu}=10^{-3}$. 
	    	
		    The results will be presented using the scaled torque\footnote{Please note that since we only use the scaled torque in this paper, the words torque and scaled-torque might be used interchangeably.}, $\Gamma/\Gamma_{0}$, where
			\begin{equation}
			  \label{eq:gamma0}
			  \Gamma_{0}= \left( \frac{q}{h} \right)^2 \, \Sigma(r_{\rm p}) \, r_{\rm p}^4 \, \Omega_{\rm p}^2 \,
			\end{equation}
			is the torque normalization which is calculated at the actual location of the planet, $r_{\rm p}$, where $q$ is the planet to star mass ratio and $\Omega_{\rm p}$ Keplerian angular velocity at $r_{\rm p}$. 

			Figure~\ref{fig:singlediscprofiles} shows the scaled torque for three disc models with the specified profiles. Each point represents the obtained torque for different surface density $\Sigma_0$ and planetary mass $M$. The torques are measured after the planet settled completely in the disc and its torque reached to a constant value. This relaxation time is 60~orbits for the zero vortensity model and 700~orbits for the other two. The theoretical torques, dotted lines, are calculated using Eq.~5,~50--53 in P11 with $P_{\chi}=0$ (meaning that we assume thermal diffusivity is infinite in a locally isothermal disc) and taking into account the effect of smoothing length from \cite{2010MNRAS.401.1950P}. In all three panels, the normalized torque on a non-migrating planet does not depend on the value of surface density or, equivalently said, on the disc mass. On the contrary, the torque on a migrating planet can increase by a factor 1.5 compared to the non-migrating value as the surface density quadruples. In the middle panel, the torque is only Lindblad and all lines for non-migrating cases almost overlap with the theoretical ones. The slightly smaller torque of the $20 M_{\oplus}$ planet is because of a very shallow partial gap around the planet. In the other two panels, the non-zero background vortensity gradient creates a contribution from the co-rotation torque that depends on the horse-shoe size and thereupon on the planet's mass. In these cases, the P11 formulae give similar values which are in agreement with the non-migrating torques with the maximum error of about 20\% for the highest planetary mass.

			Except for the model with constant background vortensity, one might wonder if the difference between the migrating and non-migrating torque can be due to the dynamical co-rotation torque \citep{2014MNRAS.444.2031P} --a possible component of the co-rotation torque that rises from the vortensity gradient created in the horse-shoe region because of the planet's migration. For example if a low-mass planet migrates a long distance in the disc while preserving its initial vortensity in the horse-shoe region, the dynamical torque can be significant. In our models, neither the planet migrates a long distance (maximum 0.24 length unit) nor is the viscosity so low to retain the initial vortensity. The conditions given in \cite{2014MNRAS.444.2031P} can help us to check whether the dynamical torque has a significant role in our simulations or not. Our setup falls into his second condition because: (a)~the variable $m_c=4\pi r_{0}^2 \Sigma_{0}/\sqrt{q h}$ (his Eq.~20) is larger than unity for our lowest planetary mass and highest surface density, (b)~our viscosity time scale over the horse-shoe region is smaller than the migration time scale $\tau_{\nu} < \tau_{mig}$, (c)~and $m_{c} \tau_{\nu} /\tau_{mig} < 1$. Therefore, the dynamical torque cannot be significant in our models. However, in long simulations for modelling of the resonant planets, this component may influence the results.

			Another point deduced from Fig.~\ref{fig:singlediscprofiles} is that the difference between the non-migrating and migrating torque is much smaller for the model with positive co-rotation torque, namely $\Sigma \propto r^{-0.5}$ and $h=0.05$. Therefore, we expect that using an inconsistent torque for this profile does not make a big difference when simulating resonant planets. We will investigate this issue in Sec.\ref{sec:parameterstudy}.

			\begin{figure}
				\centerline{\includegraphics[width=\columnwidth]{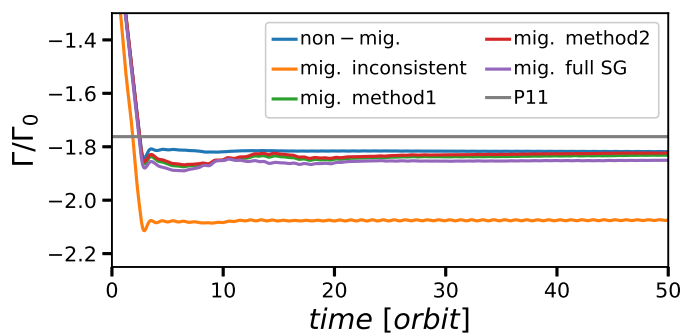}}
				\caption{Time evolution of the torque acting on a $10$~Earth-mass planet in a disc with vanishing vortensity gradient for migrating and non-migrating planets. Both of the correcting methods fix the migrating torque perfectly and return it to the non-migrating value which is identical to the full self-gravitating torque (the purple line). The horizontal line marked with P11 shows the theoretical Lindblad torque.}
				\label{fig:torquemethods}
			\end{figure}

		\subsection{Correcting the migrating torque}
		\label{subsec:correction}
\begin{figure}
			\centerline{\includegraphics[width=\columnwidth]{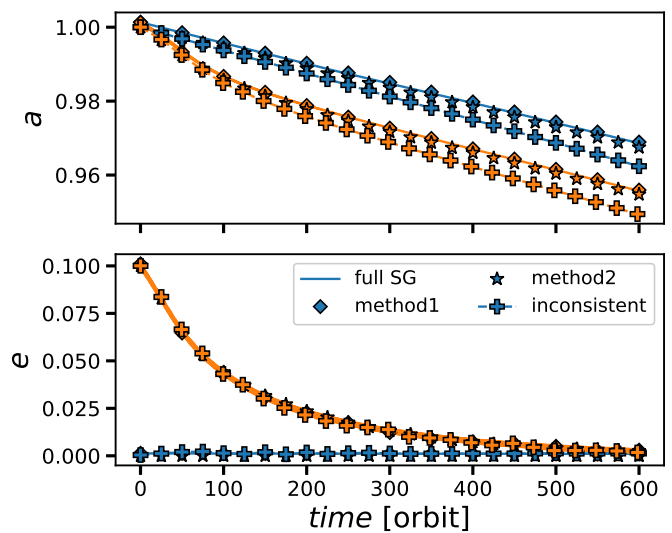}}
			\caption{Semi-major axis $a$ and eccentricity $e$ evolution of a planet with $e_0=0$ and $0.1$ {for the correct, inconsistent, and full self-gravity torques. Results of both methods excellently overlap  with the full self-gravity model}.}
			\label{fig:comparemethods}
		\end{figure}

	    	In this section, we test the two previously mentioned methods for correcting the torque (force) on a migrating planet in hydrodynamical studies and compare the results with the torque on a non-migrating planet as well as with the torque in a full self-gravitating simulation.
		For a consistent calculation of the torque, planet and gas velocities should be both computed in the same way. This can be achieved by either including or excluding the disc self-gravity simultaneously for both. Therefore, two methods are suggested in the literature:
		\begin{description}
			\item[(1)] excluding the disc self-gravity by using only the perturbed gas density for calculating the disc force on the planet. In this method, the azimuthally averaged surface density is subtracted from each cell prior to the force calculation \citep{2016ApJ...826...13B}. Therefore, the acceleration from the whole disc on the planet vanishes except where is azimuthally perturbed and the planet's velocity remains close to Keplerian. 
			
			\item[(2)] including the axi-symmetric part of the disc self-gravity as a source term in the gas momentum equation as suggested by \cite{2008ApJ...678..483B}. In this method, the gas velocities are computed using the same way as the gravitational forces on the planet. 
			
		\end{description}
		Figure~\ref{fig:torquemethods} shows the non-migrating, inconsistent migrating, corrected migrating torques, and the torque calculated taking the full self-gravity of the disc into account for a 10~Earth-mass planet in a disc with zero background vortensity and $\Sigma_0 = 2 \times 10^{-4}$. Figure~\ref{fig:comparemethods} demonstrates the migration and eccentricity damping for a circular and an eccentric planet of the same mass and disc as in Fig.~\ref{fig:torquemethods}. The symbols denote the correction methods and the inconsistent torque, while the line refers to the full self-gravity model. These two plots show that both of these methods remedy the torque perfectly and either of them must be applied in hydrodynamical simulations of multi-planets, in particular when considering the formation of resonant configurations which depends delicately on the differential migration speed of the two planets. Interestingly, eccentricity damps with the same rate in the all models.

	\section{Tighter resonances with inconsistent torque}
	\label{sec:twophydro}
		One common way to study resonant configurations is using an inner planet which is trapped or has slower migration together with an outer planet which migrates faster and catches the inner one in a resonance \citep[e.g.][]{2004A&A...414..735K,2005MNRAS.363..153P,2011MNRAS.417.2253P,2013MNRAS.434.3018P,2019ApJ...872...72C}. We use here the same approach and construct a planetary trap by increasing the disc viscosity in the inner part of the disc. Accordingly, a density maximum is created with a very steep positive density gradient that can trap the planets. Then, we planted two planets with masses $M_\mathrm{i}=10 M_{\oplus}$ and $M_\mathrm{o}=20 M_{\oplus}$ in the disc and allow them to migrate. Subscripts 'i' and 'o' refer to the inner and outer planet, respectively. Three simulations with identical initialization were run except that we used the corrected torque in one of them, the inconsistent torque in the second, and full self-gravitating calculation in the third one in order to examine what happens to the resonant configuration if one uses the inconsistent torque.
		
\begin{figure}
			\centerline{\includegraphics[width=\columnwidth]{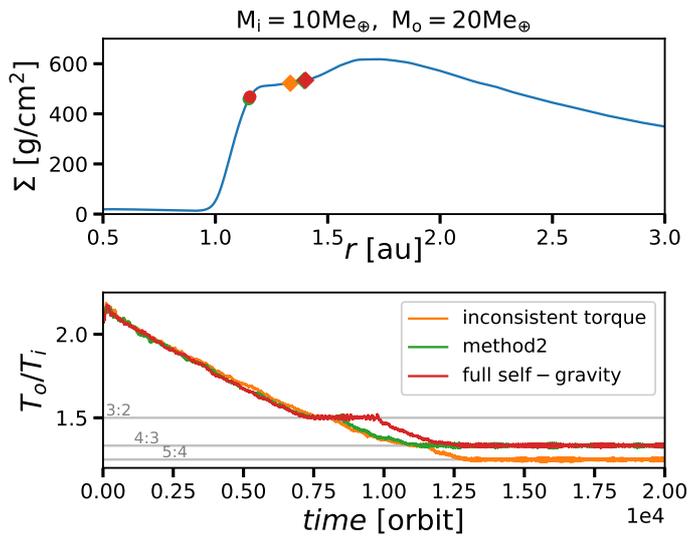}}
			\caption{Resonant capture of two planets (with 10 and 20 $M_\mathrm{Earth}$) at a planet trap located at 1au obtained by hydrodynamical simulations. Orange, green, and red colours refer to the simulations with inconsistent, corrected, and full self-gravitating torques, respectively. Top: final surface density and planet positions. Bottom: Time evolution of the period ratio (outer to inner planet) of the two planets.
          }
			\label{fig:twoplanets}
		\end{figure}

		The upper panel in Fig~\ref{fig:twoplanets} shows the evolved surface density and the positions of the planets after the planets have been locked into resonance. This profile is almost time-independent and very similar in all three simulations. The lower panel of Fig~\ref{fig:twoplanets} compares the orbital period ratio of the outer to inner planet between the models. The final location of these planets are also marked in the upper panel with the corresponding colours. This figure clearly signifies the role of correction. While the inner planet is trapped in all models at the zero torque location \citep{2006ApJ...642..478M}, the outer planet migrates more inwards in the model with inconsistent torque than the one with the correct torque, which produced the same final configuration as the model with full self-gravity. This can be explained easily using the results in Sec.~\ref{sec:torquesingle}. Because the inconsistent torque is larger than the correct one, even for the low surface densities as we have here, the planet migrates faster and crosses the wider resonances. This is why the outer planet in the model with inconsistent torque was able to pass through more resonances until it finally reached the 5:4 commensurability. On the other hand, the model with correct torque has been stopped in the wider 4:3 resonance. As a result, an incorrect torque treatment (i.e. not considering disk self-gravity at all)
will lead to capture in closer resonances, which might impact the stability of the system.
		
	\section{Under what condition can using inconsistent torque be troublesome?}
	\label{sec:parameterstudy}
	Because the migration rate of a planet depends directly on its mass and the disc surface density, we investigate in this section under what conditions ignoring the torque (resp. self-gravity) correction can be risky. Although the migration rate depends inversely on the disc aspect ratio, we will not study this parameter because for small aspect ratios, even moderate planetary masses can open a partial gap and enter into a different migration regime which is not within the scope of this study.
		A parameter study is needed to find out for what values of disc surface densities and planetary masses the final resonance configuration depends strongly on the torque correction. However, performing such a parameter study using hydrodynamical simulation would be numerically expensive. To circumvent this issue, we use N-body simulations which are much faster. However, we first need to know how much the results of N-body simulations agree with hydro simulations for correct and inconsistent torque.
		
		\subsection{N-body versus hydro simulations}
		\label{subsec:nbodyvshydro}
		Among the models in Sec.~\ref{subsec:lowmass}, the disc with $\Sigma = \Sigma_{\rm 0} (r/r_{\rm 0})^{-0.5}$ and constant aspect ratio shows the least difference between the migrating and non-migrating torques. Therefore, this model can be considered as the most conservative one. If we see considerable differences in the final configurations, the case for other disc models would be even worse. Henceforth, we consider this model with $\Sigma_{0}=2\times10^{-4}$, corresponding to $1777\rm [g/cm^2]$ for a solar-mass star and the length unit of 1~au, and $h=0.05$. We place two planets with masses $M_\mathrm{i}=M_\mathrm{o}=10M_{\oplus}$ just out of 2:1 resonance and allow the system to evolve. The setup for the hydro simulations is otherwise identical to Sec.~\ref{subsec:numericshydro}. 
		
		The N-body simulations have been carried out using the \texttt{REBOUND} code\footnote{Available at http://github.com/hannorein/rebound.} \citep{2012A&A...537A.128R} with the IAS15 integrator \citep{2015MNRAS.446.1424R}, and identical planet initialisation as our hydro simulations. Migration of the planets is modelled using the torque formulae in P11 along with the correction factors for the planet's eccentricity as in \cite{2008A&A...482..677C} for Lindblad and \cite{2014MNRAS.437...96F} for co-rotation torque. Converting the torques to the planet acceleration has been done as in \cite{2011CeMDA.111...83P}. Because we also want to simulate models with inconsistent torque, inspired by \cite{2008ApJ...678..483B}, we fitted a function to the data in Sec.~\ref{subsec:lowmass} that gives us the ratio of the {\it inconsistent} torques $\Gamma_{\rm mig}$ to the non-migrating ones $\Gamma_{\rm fix}$. This function reads
		\begin{equation}
			\label{eq:fix}
			\frac{\Gamma_{\rm mig}}{\Gamma_{\rm fix}} = 1+ 5.23 Q^{-1.04},
		\end{equation}
		where $Q$ is the Toomre parameter at the location of the planet\footnote{Note that the  fitting in \cite{2008ApJ...678..483B} has been done over $Qh$. In this study, since we do not vary the value of $h$ at the location of the planet, it has been enclosed in the coefficient.}. Therefore, for modelling the inconsistent torque in the N-body simulations, we apply this relation on the total torque from P11.	
		
		For the disc and the planetary masses used here, we expect that the planets keep their initial period ratio without diverging or converging. When two inwardly migrating planets converge, the outer planet has a larger migration rate than the inner. In other words, $\dot{a}_\mathrm{o} > \dot{a}_\mathrm{i}$ with $a$ being the planet's semi-major axis. The migration rate of a non-eccentric planet is related to the torque as
		\begin{equation}
		\label{eq:tmig}
			\dot{a} = \frac{2\Gamma a^{1/2}}{q},
		\end{equation}
		where the torque $\Gamma$ is determined by (a)~the disc surface density and temperature slopes $-\alpha$ and $-\beta$ which are constant in our models, and (b)~$\Gamma_{0} \propto q^2 a^{-\alpha+1-2f}$ (see Eq.~(\ref{eq:gamma0})), that varies as the planet migrate. Therefore, the latter determines if the migration of our planets diverges or converges. Applying the condition of viscous equilibrium gives $\Gamma_{0} \propto q^2 a^{-2\alpha+3/2}$, and substituting $\Gamma_{0}$ in Eq.~\ref{eq:tmig} gives
		\begin{equation}
		\label{eq:tmigdepr}
			\dot{a} \propto {q a^{-2\alpha+1}}. 
		\end{equation}
		The convergence condition then reads
		\begin{equation}
		\label{eq:convergecondition}
			\frac{\dot{a}_\mathrm{o}}{\dot{a}_\mathrm{i}} = \frac{q_\mathrm{o}}{q_\mathrm{i}} \left(\frac{a_\mathrm{o}}{a_\mathrm{i}}\right)^{-2\alpha+1} > 1.
		\end{equation}
		For equal mass planets, this condition is satisfied only if $\alpha>0.5$. Therefore, for our setup, both migration time scales are equal and, on paper, we expect they neither converge nor diverge. However, the condition in a hydro simulation would be different due to the planet-spiral \citep{2013ApJ...778....7B} or planet-planet interactions .
		
\begin{figure}
			\centerline{\includegraphics[width=\columnwidth]{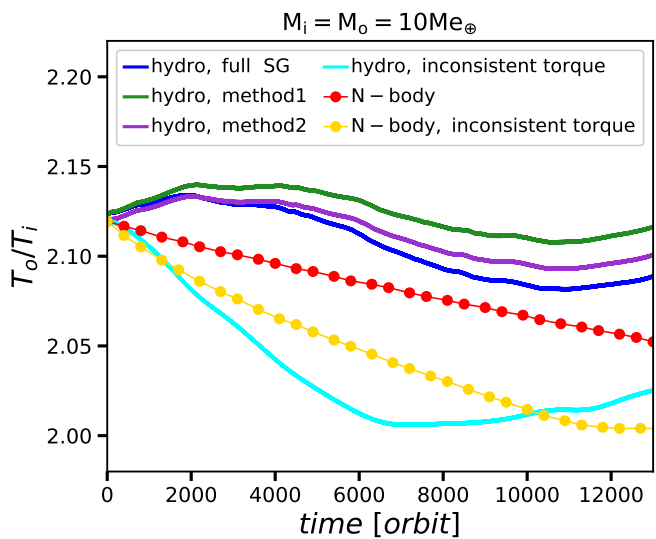}}
			\caption{Orbital period ratio of two equal mass planets in a disc with $\Sigma \propto r^{-0.5}$ and $h=0.05$ using an N-body and a hydro code. The inconsistent torques in the N-body simulation has been computed based on the data in Sec.~\ref{subsec:lowmass}, see Eq.~(\ref{eq:fix}).}
			\label{fig:hdrovsnbody}
		\end{figure}

		Figure~\ref{fig:hdrovsnbody} shows the orbital period ratio of our planets for four hydro and three N-body simulations, all of them initialised identically. Three hydro simulations in which we used either the correct torque or ran with full self-gravity agree well and do not show a considerable divergence or convergence during the running time. In contrast, the hydro simulation with the inconsistent torque diverges into 2:1 resonance. This is the result that is also produced by the N-body simulation where we used Eq.~\ref{eq:fix} to mimic the inconsistent torques. Differently than the correct hydro simulations, the N-body simulation with the correct torque shows convergence. This behaviour happens when we apply the eccentricity correction on the Lindblad torques. A slight eccentricity of the outer planet increases the torque such that they converge by time. Although the N-body gives convergent migration compared to the hydro simulation, it is slower than the one with the inconsistent torque and we can still use the N-body simulations for the parameter study, keeping in mind that the results underestimate the difference between the models with inconsistent and the correct torque. If we see a notable difference, it would be even larger when using hydro simulations.

		\subsection{Parameter study}
		\label{subsec:rebound}
\begin{figure}
			\centerline{\includegraphics[width=\columnwidth]{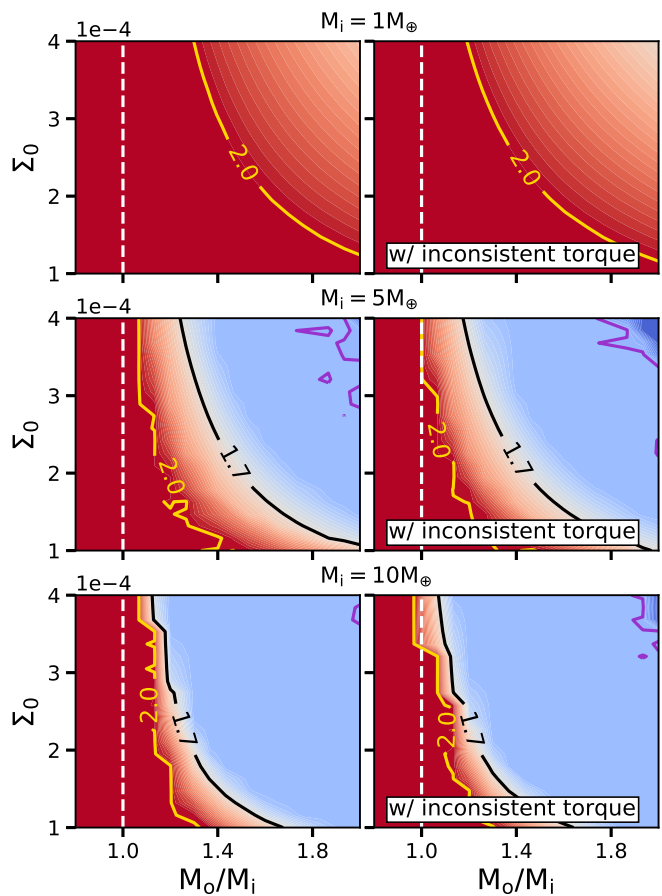}}
			\caption{Orbital period ratio of the outer to inner planet for models with correct torque (left column) and the ones with inconsistent torque (right column) as a function of planet mass ratio and surface density. Some commensurabilities are marked by coloured lines: 2:1 by yellow, 5:3 by black, and 3:2 by purple. Visibly, using inconsistent torque can produce more packed configurations.}
			\label{fig:tratiomaps}
		\end{figure}

		Inspired by the \textit{Kepler} data in Fig.~\ref{fig:keplermasses}, we focus on planets with masses between $[1\textup{--}20]M_{\oplus}$ and an outer-to-inner planet mass ratios between $[0.8\textup{--}2]$. Surface density is varied as $\Sigma_{0}\in[1\textup{--}4]\times 10^{-4}$, from half to twice of the MMSN's value. We initialized the system as in Sec.~\ref{subsec:nbodyvshydro} and allowed it to evolve for 10\,000~years using the N-body code. The results are summarised in Fig.~\ref{fig:tratiomaps}. The left panels, which display the results of the simulations with correct torques, show that all models with mass ratios below 1.1 do not converge, regardless of the surface density values. Only when the outer planet becomes more massive than the inner, the migration is converging. On the contrary, the models with inconsistent torques, panels on the right, can produce more packed systems. This can be understood by comparing the location of the marked contours, which show some commensurabilities. For example, equal mass planets reach the 2:1 resonance for higher surface densities while the correct torque simulations do not predict it. Considering the N-body results favour smaller difference between the two sets, we can conclude that applying the torque correction in hydro simulations would produce wider configurations.
		
	\section{Summary}
	\label{sec:summary}
		Migration of planets is the result of gravitational interaction between the planets and their natal disc. A planet plays the role of a perturber in a disc and creates spiral arms as the consequence of its interaction with the gas at Lindblad resonances. Location of these resonances is a function of relative angular velocity of the gas to the planet, namely $\Omega(r)-\Omega(r_{\rm p})$. In order to calculate the planetary migration correctly, the computation of these two velocities should be consistent, meaning that the disc gravity should be either included or excluded in calculating the force acting on both, the gas and planet. However, this is often ignored in hydrodynamical simulations when the disc mass is low, and the disc force only acts on the planet. We showed here that, due to this inconsistent calculation, a slightly larger torque can make more packed planetary systems and even fake resonant configurations. We examined two suggested methods in the literature for correcting the torque: 
		\begin{description}
			\item[(1)]using only the surface density perturbation for calculating the force from disc on the planet while ignoring the disc self-gravity \citep{2016ApJ...826...13B,2008ApJ...678..483B}. In this method, the disc gravity, except from the perturbed gas, is excluded from calculations.
			\item[(2)]including the axisymmetric part of the disc self-gravity in the gas momentum equation \citep{2008ApJ...678..483B}. In this way, the disc gravity is included in calculation of both planet and gas. For a convenient numerical procedure see also \citet{1996MNRAS.282..234K}. 
		\end{description}
		We showed that both of these methods give identical results as the models with full self-gravity. Considering that they take similar computational time, one of these corrections must be applied in hydrodynamical modelling of planetary systems, otherwise, the results would not be very reliable. Here, the inclusion of axisymmetric disc self-gravity is physically more realistic and can be generalized to full self-gravity if required.
		
		We examined this issue for low-mass planets in the mass range of \textit{Kepler} planets through customized N-body simulations. For this we used the most conservative disc setup with a density slope for which the differences between consistent and inconsistent forces were minimal, hence we expect an even stronger effect for other situations. Since the base of the calculations and the physics is the same for higher mass planets, we expect that this issue affects the migration of higher mass planets as well, may be less pronounced due to the their gap opening.
		
		We would like to emphasise that the calculation of popular torque formulae, which are widely used in N-body simulations, are consistent because in those studies the planet is enforced to move Keplerian while the gas self-gravity is ignored. Such a condition, as we explain before, is consistent. There is also another method of calculation in the literature that the planet is forced to migrate with a given migration rate and then the torque on the planet is measured  \citep{2014ApJ...792L..10D,2019MNRAS.482.3678M}. We found that this method is also consistent since the planet's velocity is not calculated directly from the gas but is set actually by hand. Our warning refers mostly to the models in which the planet's migration is calculated directly through its gravitational interaction by the disc.

		Finally, we would like to point out that only full hydrodynamical simulations including self-gravity corrections will give reliable results as additional effects caused by gravitational interaction with the spiral arms produced by the planets will play a role. These are difficult to capture using N-body simulations.
	\begin{acknowledgements}
		We acknowledge the support of the DFG priority program SPP 1992 "Exploring the Diversity of Extrasolar Planets under grant KL 650/27". The authors also acknowledge support by the state of Baden-Württemberg through bwHPC. We would like to thank A.~Crida for the stimulating discussion and the anonymous referee for her/his comments which helped improving the study.
	\end{acknowledgements}

	\bibliographystyle{aa}
	\bibliography{sgmatters}
	
\end{document}